\documentclass[times,5p,preprint]{elsarticle}
\pdfoutput=1 

\usepackage[utf8]{inputenc}
\usepackage{hyperref}
\usepackage{url}
\graphicspath{{./img/}}
\usepackage{tikz}
\usetikzlibrary{arrows,shadows}
\usepackage{pgf-umlsd}

\usepackage{amssymb}
\usepackage{amsmath}
\usepackage{nicefrac} 

\usepackage{tabularx}
\usepackage{tabulary}
\usepackage{booktabs}
\usepackage{array}
\usepackage{mdwmath}
\usepackage{eqparbox}
\usepackage[caption=false,font=footnotesize]{subfig}

\usepackage{fixltx2e}

\usepackage[tracking=true]{microtype}
\usepackage{ellipsis}

\usepackage{listings}
\makeatletter
\providecommand*{\toclevel@lstlisting}{0}
\makeatother

\usepackage{svg}

\newcommand{\executeiffilenewer}[3]{%
 \ifnum\pdfstrcmp{\pdffilemoddate{#1}}%
 {\pdffilemoddate{#2}}>0%
 {\immediate\write18{#3}}\fi%
}
\newcommand{\includesvgnolatex}[2][1]{%
 \executeiffilenewer{#2.svg}{#2.pdf}%
 {inkscape -z -D --file=#2.svg --export-pdf=#2.pdf}%
 \includegraphics{#2.pdf}
}

\hypersetup{
  pdftitle={DiaSys: Improving SoC Insight Through On-Chip Diagnosis},
  pdfauthor={Philipp Wagner, Thomas Wild, and Andreas Herkersdorf},
}

\journal{Journal of Systems Architecture}

\begin{document}
\begin{frontmatter}

\title{DiaSys: Improving SoC Insight Through On-Chip Diagnosis}

\author[tumaddress]{Philipp Wagner\corref{mycorrespondingauthor}}
\ead{philipp.wagner@tum.de}
\author[tumaddress]{Thomas Wild}
\ead{thomas.wild@tum.de}
\author[tumaddress]{Andreas Herkersdorf}
\ead{herkersdorf@tum.de}

\cortext[mycorrespondingauthor]{Corresponding author}
\address[tumaddress]{Institute for Integrated Systems, Technical University of Munich, Arcisstraße 21, 80333 Munich, Germany}

\begin{abstract}
To find the cause of a functional or non-functional defect (bug) in software running on a multi-processor System-on-Chip (MPSoC), developers need insight into the chip.
Tracing systems provide this insight non-intrusively, at the cost of high off-chip bandwidth requirements.
This I/O bottleneck limits the observability, a problem becoming more severe as more functionality is integrated on-chip.
In this paper, we present DiaSys, an MPSoC diagnosis system with the potential to replace today's tracing systems.
Its main idea is to partially execute the analysis of observation data on the chip; in consequence, more information and less data is sent to the attached host PC.
With DiaSys, the data analysis is performed by the diagnosis application.
Its input are events, which are generated by observation hardware at interesting points in the program execution (like a function call).
Its outputs are events with higher information density.
The event transformation is modeled as dataflow application.
For execution, it is mapped in part to dedicated and distributed on-chip components, and in part to the host PC; the off-chip boundary is transparent to the developer of the diagnosis application.
We implement DiaSys as extension to an existing SoC with four tiles and a mesh network running on an FPGA platform.
Two usage examples confirm that DiaSys is flexible enough to replace a tracing system, while significantly lowering the off-chip bandwidth requirements.
In our examples, the debugging of a race-condition bug, and the creation of a lock contention profile, we see a reduction of trace bandwidth of more than three orders of magnitude, compared to a full trace created by a common tracing system.
\end{abstract}


\end{frontmatter}

\section{Introduction}
To write high-quality program code for a Multi-Processor System-on-Chip (MPSoC), software developers must fully understand how their code will be executed on-chip.
Debugging and tracing tools can help developers to gain this understanding.
They are a keyhole through which developers can peek and observe the software execution.
Today, and even more in the future, this keyhole narrows as MPSoCs integrate more functionalities, while at the same time the amount of software increases dramatically.
Furthermore, concurrency and deep interaction of software with hardware components beyond the instruction set architecture (ISA) boundary are on the rise.
Therefore more, not less, insight into the system is needed to keep up or even increase developer productivity.

Many of today's MPSoCs are executing concurrent code on multiple cores, interact with the physical environment (cyber-physical systems), or must finish execution in a bounded amount of time (hard real-time).
In these scenarios, a non-intrusive observation of the software execution is required, like it is provided by tracing.
Instead of stopping the system for observation, as done in run-control debugging, the observed data is transferred off-chip for analysis.
Unfortunately, observing a full system execution would generate data streams in the range of petabits per second~\cite[p.~16]{vermeulen_debugging_2014}.
This is the most significant drawback of tracing: the system insight is limited by the off-chip bottleneck.


Today's tracing systems, like ARM CoreSight~\cite{CoreSightWebsite} or NEXUS 5001~\cite{_nexus_2003} are designed to efficiently capture the operation of a functional unit (like a CPU) as compressed trace stream.
With filters and triggers it is possible to configure which and when a functional unit is traced (observed).
The trace streams (or short, traces) are then transported across an off-chip interface (and possibly other intermediate devices) to a host PC.
Upon arrival the compressed trace streams are first decompressed (reconstructed) using the program binary and other static information which was removed before.
The reconstructed trace streams are then fed to a data analysis application, which extracts information out of the data.
This information can then be presented to a developer or it can be used by other tools, e.g. for runtime verification.
\medskip

The \textbf{main idea} in this work is to move the data analysis (at least partially) from the host PC into the chip.
Bringing the computation closer to the data sources reduces the off-chip bandwidth requirements, and ultimately increases insight into software execution.

To realize this idea, we introduce \emph{DiaSys}, a replacement for the tracing system in an MPSoC.
DiaSys does not stream full execution traces off-chip for analysis.
Instead, it first creates events from observations on the chip.
Events can signal any interesting state change of the observed system, like the execution of a function in the program code, a change in interconnect load beyond a threshold, or the read of a data word from a certain memory address.
A \emph{diagnosis application} then processes the observed events to give them ``meaning.''
Given an appropriate diagnosis application, a software developer might not be presented with any more a sequence of events like ``a read/write request was issued'', but with the more meaningful output of the diagnosis application ``a race condition bug was observed.''
Analyzing the data on-chip is not only beneficial to reduce the off-chip bandwidth requirements, but also enables new use cases in the future, such as self-adapting or self-healing systems.

However, doing all this processing on-chip would, in some cases (and markets), be too costly in terms of chip area.
Therefore, we describe the diagnosis applications so that they can be transparently split into multiple parts: one part executing on-chip in dedicated, distributed hardware components close to the data source, and another part running on a host PC.

In summary, our \textbf{key contributions} are:
\begin{itemize}
 \item an architecture and component library of on-chip infrastructure to collect and analyze diagnosis data created during the software execution, and
 \item a model of computation which allows developers to describe data analysis tasks (the ``diagnosis application'') in a way which is independent of the specific hardware implementation of the diagnosis system.
\end{itemize}

Combining these two contributions, we show that
\begin{itemize}
 \item DiaSys is a viable alternative to tracing systems in the two major fields where tracing is employed today: hypothesis testing (debugging) and the collection of runtime statistics. Two case studies explore these use cases (Section~\ref{sec:usage}).
 \item the diagnosis applications introduced by DiaSys are a beneficial representation of a data analysis task: they abstract from the implementation through a clearly defined model of computation to foster re-use and portability (Section~\ref{sec:method:diagnosis_applications}).
 \item DiaSys is implementable in hardware with reasonable system cost (Section~\ref{sec:hwimpl}).
\end{itemize}

In the following, we explore our diagnosis system in depth.
We start with a thorough analysis of the state of the art in tracing systems in Section~\ref{sec:related_work},
based on which we developed our concept of the diagnosis system presented in Section~\ref{sec:method}.
We include a detailed model of a diagnosis application and a discussion of its semantics in Section~\ref{sec:method:diagnosis_applications}.
The architecture of our diagnosis system is presented next in Section~\ref{sec:arch}, followed by a discussion of its possible limitations.
In Section~\ref{sec:hwimpl} we present our hardware implementation.
Combining all parts, we show two usage examples in Section~\ref{sec:usage}, one to find a multi-core race condition bug, and one to create an application profile.


\paragraph{A word on terminology} We use the terms ``diagnosis'' and ``diagnosis system'' to stress the fact that we integrate the on-chip observation of the software execution with the data analysis.
We use ``tracing'' to refer to the method of obtaining insight into the SoC by transferring a stream of observations from a functional unit off-chip.
``Debugging'' is used as a synonym for ``hypothesis testing,'' the process of (usually manually) checking (by various means) if the software behaves as expected.

\section{Background and Related Work}
\label{sec:related_work}

Our approach touches and integrates two usually separated topics: obtaining a software execution trace from a SoC, and processing the obtained information in order to generate useful information.
In this section we present background and related work on both topics.

\subsection{Gaining Insight into SoCs}
\label{sec:related_work:insight}
\begin{figure}
 \centering
 \includesvgnolatex{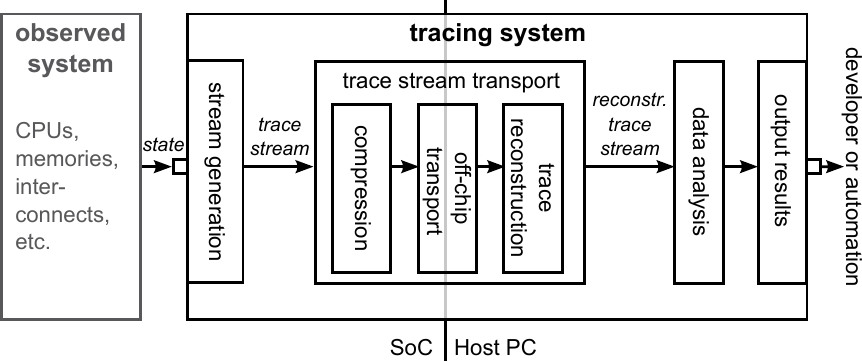}

 \caption{A schematic view of a common tracing system like ARM CoreSight or NEXUS 5001.}
 \label{fig:diagnosis_applications_existing}
\end{figure}

Today's tracing solutions for SoCs are designed to capture and transfer as much as possible of the SoC's internal state to an external observer.
They are generally structured as shown in Figure~\ref{fig:diagnosis_applications_existing}.

First, trace data streams are obtained from the observation of various functional units in the system, like CPUs, buses and memories.
Then, this data is spatially and temporally reduced through the use of filters and triggers.
Finally, the redundancy in the data is removed by the use of compression algorithms.
The resulting trace data stream is then transferred off-chip (live or delayed through an on-chip memory buffer).
On a host PC, the original trace streams are reconstructed using information from the program binary and other static information, which was discarded as part of the compression process.

All major commercial SoC vendors offer tracing solutions based on this template.
ARM provides its licensees the CoreSight intellectual property (IP) blocks~\cite{CoreSightWebsite}.
They are used in SoCs from Texas Instruments, Samsung and STMicroelectronics, among others.
Vendors such as Qualcomm (formerly Freescale) include tracing solutions based on the IEEE-ISTO~5001 (Nexus) standard~\cite{_nexus_2003}, while Infineon integrates the Multi-Core Debug Solution (MCDS) into its automotive microcontrollers~\cite{ipextreme_infineon_2008}.
Since 2015 Intel also includes a tracing solution in their desktop, server and embedded processors called Intel Processor Trace (PT)~\cite{_intel_2015}.
The main differentiator between the solutions is the configurability of the filter and trigger blocks.

Driven by the off-chip bottleneck, a major research focus are lossless trace compression schemes.
Program trace compression available in commercial solutions typically requires 1 to 4~bit per executed instruction~\cite{hopkins_debug_2006, orme_debug_2008}, while solutions proposed in academia claim compression ratios down to $0.036$~bit per instruction~\cite{uzelac_real-time_2010}.
Even though data traces contain in general no redundancy, in practice compression rates of about 4:1 have been achieved \cite{hopkins_debug_2006}.

\subsection{Analyzing System Behavior}
\label{sec:related_work:analysis}
A human is easily overwhelmed when asked to analyze multiple gigabits of trace data each second.
Instead, automated analysis tools are used to extract useful information out of the vast amount of trace data.
Such tools have one common goal: to help a developer better understand the software execution on the target system.
The means to achieve this goal, however, vary widely.

Diagnosis applications which analyze non-functional issues such as performance bugs often generate results in the form of ordered lists.
They list for example applications which consume most processing or memory resources, or which generate most I/O traffic.
This report can then be a starting point for a more fine-grained analysis of the problem.
Diagnosis applications which target functional bugs are usually more specialized; in many cases, a diagnosis application is created just to confirm or negate one single hypothesis about the software execution on the chip.
For example, a developer might want to confirm that a certain variable stays within defined bounds, e.g. to check if an array overflow occurred.

Most analysis tools for SoCs are not stand-alone applications, but part of debugging and tracing software packages from vendors like Lauterbach, Green Hills or ARM.
They are usually controlled through a graphical user interface.

Of course, analysis applications used to understand software execution are not only developed for SoCs and other embedded systems.
Most tools in this domains are intrusive: they run as part of the analyzed system and obtain the required system state through instrumentation.
However, the general concepts are also relevant for the diagnosis of SoCs.
This is especially true for scriptable or programmable debugging, which applies the concept of event-driven programming to debugging.
Whenever a defined \emph{probe point} is hit, an event is triggered and an \emph{event handler} executes.
Common probe points are the execution of a specific part of the program (like entering a certain program function), or the access to a given memory location.
The best-known current implementations of this concept are DTrace and SystemTap, which run on, or are part of, BSDs, Linux, and macOS (where DTrace is integrated into the ``Apple Instruments'' product) \cite{cantrill_dynamic_2004,eigler_architecture_2005}.
The concept, however, is much older.
Dalek~\cite{olsson_dataflow_1991} is built on top of the GNU Debugger (GDB) and uses a dataflow approach to combine events and generate higher-level events out of primitive events.
Marceau et al. extend the dataflow approach and apply it to the debugging of Java applications~\cite{marceau_dataflow_2004}. Coca~\cite{ducasse_coca_1999}, on the other hand, uses a language based on Prolog to define conditional breakpoints as a sequence of events described through predicates for debugging C programs.
In a work targeting early multi-processor systems, but otherwise closely related to our approach, Lumpp et. al. present a debugging system which is based on an event/action model~\cite{lumpp_specification_1990}.
A specification language is used to describe events in the system trigger which an action, and hardware units can be used to identify these events.

None of the presented works directly tackle the observability problem in SoCs by moving the data analysis partially on-chip.
However, they form a strong foundation of ideas, which inspired us in the design of the diagnosis system. It is presented in the following sections.

\section{DiaSys, Our Diagnosis System}
\label{sec:method}

We have designed our diagnosis system to address the shortcomings of today's tracing systems.
Based on a set of requirements, we discuss the design of the diagnosis system in depth, followed by a hardware/software architecture implementing the diagnosis system.
First, however, we define some terms used in the following discussion.

\subsection{Definitions}
\paragraph{Observed system} The part of the SoC which is observed or monitored by the diagnosis system.
In other works, the term ``target system'' is used.

\paragraph{Functional unit} A subset of the observed system which forms a logical unit to provide a certain functionality.
Examples for functional units are CPUs, memories, or interconnect resources such as a bus or NoC routers.

\paragraph{State} The state of a system is the unity of all stored information in that system at a given point in time which is necessary to explain its future behavior.~\cite[p.~103]{harris_digital_2012} In a sequential circuit, the state is equal to the memory contents of the system.


\subsection{Design Requirements for the Diagnosis System}
\label{sec:method:requirements}
A set of requirements guides the design of the diagnosis system.

 \paragraph{Distributed} The diagnosis system must be able to reduce the amount of observation data as close to the source, i.e. the functional units, as possible.
Since the data sources are distributed across the chip, the diagnosis system must also be distributed appropriately.

\paragraph{Non-Intrusive} The diagnosis system must be non-intrusive (passive).
Non-intrusive observation preserves the event ordering and temporal relationships in concurrent executions, a requirement for debugging multi-core, real-time, or cyber-physical systems~\cite{fidge_fundamentals_1996}.
Non-intrusiveness also gives a developer the confidence that he or she is observing a bug in the program code, not chasing a problem caused by the observation (a phenomenon often called ``Heisenbug''~\cite{gray_why_1986}).

\paragraph{Flexible On-Chip/Off-Chip Cost Split} The diagnosis system must be flexible to implement.
The implementation of the diagnosis system involves a trade-off between the provided level of observability and the system cost.
The two main cost contributions are the off-chip interface and the chip area spent on diagnosis extensions.
The diagnosis system concept must be flexible enough to give the chip designer the freedom to configure the amount of chip resources, the off-chip bandwidth and the pin count in a way that fits the chip's target market.
At the same time, to provide flexibility on the observation, the system must be able to adapt to a wide range of bugs.

\paragraph{Relaxed Timing Constraints} The diagnosis system must not assume a defined timing relationship between the individual distributed components.
Today's larger SoCs are designed as globally asynchronous, locally synchronous (GALS) systems with different power and clock domains, where no fixed time relationship between components can be given.

\subsection{The Concept of the Diagnosis System}
\label{sec:method:concept}
\begin{figure}
 \centering
 \includesvgnolatex{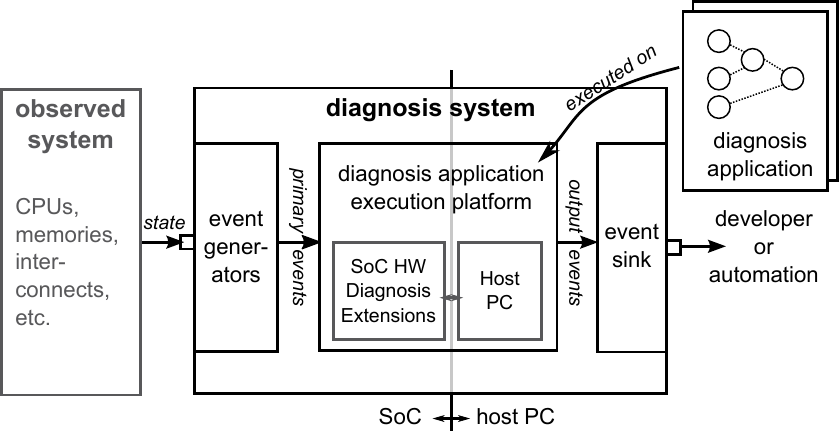}
 \caption{A schematic view of the diagnosis system.}
 \label{fig:diasys_model}
\end{figure}

Based on the discussed requirements this section gives an overview on the diagnosis system as shown in Figure~\ref{fig:diasys_model}.

The \textit{input} to the diagnosis system is the state of the observed system over time, the \textit{output} are the diagnosis results, which can be represented in various forms.
With respect to the input and output interfaces, the diagnosis system is identical to a traditional tracing system.
The difference lies in the components which generate the output from the input.
Three main components are responsible for this processing: the event generators, the diagnosis application together with its execution platform, and the event sinks. Between these components, data is exchanged as diagnosis events.

\textit{Diagnosis events are the container for data} exchanged in the diagnosis system.
In the general case, an event consists of a type identifier and a payload.
Events are self-contained, i.e. they can be decoded without the help from previous or subsequent events.

\textit{Event generators produce} primary events based on state changes in the observed system.
They continuously compare the state of the observed system with a \emph{trigger condition}.
If this condition holds, they \emph{trigger} the generation of a primary event.

A \emph{primary event} is a specialized diagnosis event in which the type identifier is equal to a unique identifier describing the event generator.
The payload contains a partial snapshot of the state of the observed system at the same instant in time as the event was triggered.
Which parts of the state are attached to the event is specified by the event generator.
For example, a CPU event generator might produce primary events when it observes a function call and attach the current value of a CPU register as payload.
A primary event answers two questions:
why was the event generated, and in which state was the observed system at this moment in time.

\textit{The diagnosis application analyzes} the software execution on the observed system.
It is modeled as transformational dataflow application, which transforms primary events into output events.
The goal of this transformation is to interpret the state changes represented in primary events in a way that yields useful information for a developer or an automated tool.
We describe diagnosis applications in more detail in Section~\ref{sec:method:diagnosis_applications}.

\textit{The diagnosis application execution platform} executes diagnosis applications.
The execution platform can span (transparent to the diagnosis application developer) across the chip boundary.
On the chip, it consists of specialized hardware blocks which are able to execute (parts of) the diagnosis application.
On the host PC, a software runtime environment enables execution of the remaining parts of the diagnosis application.
The on- and off-chip part of the execution platform are connected through the off-chip interface.
This split design of the execution platform allows hardware designers to trade off chip area with the bandwidth provided for the off-chip interface, while retaining the same level of processing power, and in consequence, system observability.

\textit{Event sinks consume} output events produced by the diagnosis application.
Their purpose is to present the data either to a human user in a suitable form (e.g. as a simple log of events, or as visualization), or to format the events in a way that makes them suitable for consumption by an automated tool, or possibly even for usage by an on-chip component.
An example usage scenario for an automated off-chip user is runtime validation, in which data collected during the runtime of the program is used to verify properties of the software.

Together, event generators, the diagnosis application and the event sink build a processing chain which provides a powerful way to distill information out of observations in the SoC.

\subsection{Diagnosis Applications}
\label{sec:method:diagnosis_applications}

Diagnosis applications are the heart of the diagnosis system, as they perform the ``actual work'' of interpreting what happens on the observed system during the software execution.
Diagnosis applications are transformational dataflow applications.
We chose this model to enable the transparent mapping of the diagnosis application to an execution platform spanning across the chip boundary.
Our goal is that the developer of the diagnosis application does not need to explicitly partition the diagnosis application into an on-chip and an off-chip part; instead, this mapping could be performed in an automated way. (Currently, however, we do not perform an automated mapping.)
No matter how the diagnosis application is mapped onto the execution platform, the behavior of the application follows identical rules, i.e. the semantics of the application stay the same.

The diagnosis application is a \emph{transformational application}, in contrast to reactive or interactive applications~\cite{halbwachs_synchronous_1991}.
This means, starting from a given set of inputs, the application \emph{eventually} produces an output.
The application code only describes the functional relationship between the input and the output, not the timing when the output is generated.
The application also does not influence or interact in another way with the observed system from which its inputs are derived.

The diagnosis application is structured as \emph{dataflow application}.
Its computation is represented by a directed graph, in which the nodes model the computation, and the edges model communication links.
In diagnosis applications we call the graph nodes \emph{transformation actors}, and the graph edges \emph{channels}.

\begin{figure}
 \centering
 \includesvgnolatex{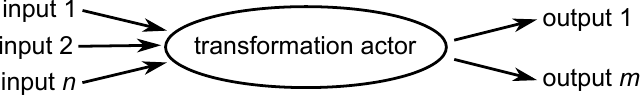}
 \caption{A transformation actor with $n=3$ input channels and $m = 2$ output channels.}
 \label{fig:transformation_actor}
\end{figure}

Each transformation actor reads events from $n \in \mathbb{N}_0$ input channels, and writes events to $m \in \mathbb{N}_0$ output channels, as shown in Figure~\ref{fig:transformation_actor}.

A transformation actor starts its processing, it ``fires,'' if a \emph{sufficient} number of events are available at its inputs.
The definition of ``sufficient'' depends on the individual transformation actor.
For example, one transformation actor might always read one event from each input before starting the processing, while another one might always read two events from input 1 and one event from input 2.

When firing, the transformation actor applies an arbitrary transformation function $f$ to the input events.
The generated output depends on
\begin{itemize}
 \item the read input events,
 \item the ordering of the input events, and
 \item the internal state of the transformation actor.
\end{itemize}
Transformation actors may communicate only through the input and output channels, but not through additional side channels (e.g. shared variables).

Diagnosis applications built out of such transformation actors are \emph{nondeterministic}, as defined by Kahn~\cite{kahn_semantics_1974,lee_dataflow_1995}.
This means, the output not only depends on the history of inputs (i.e. the current input and the state of the actor), but also on the relative timing (the ordering) of events.

Nondeterministic behavior of diagnosis applications is, in most cases, the expected and wanted behavior; it gives its authors much needed flexibility.
An example of nondeterministic diagnosis applications are applications which aggregate data over time, like the lock contention profiling presented in Section~\ref{sec:usage:lockprofiling}.
These applications consume an unspecified amount of input events and store an aggregate of these inputs.
After a certain amount of time, they send a summary of the observations to an output channel.

But at the same time, nondeterministic diagnosis applications prevent the static analysis of event rates, bandwidth and processing requirements.
If wanted, application authors can therefore create deterministic diagnosis applications, if they restrict themselves to
\begin{itemize}
 \item always reading the input channels in the same order without testing for data availability first (instead, block and wait until the data arrives),
 \item connecting one channel to exactly one input and one output of an actor, and
 \item using only transformation functions which are deterministic themselves.
\end{itemize}

Note that we only describe the diagnosis application itself as nondeterministic.
Its execution, after being mapped to an execution platform, can be deterministic, i.e. multiple identical runs produce the same diagnosis result.

\subsection{Diagnosis System Architecture}
\label{sec:arch}
\begin{figure}
 \includesvgnolatex{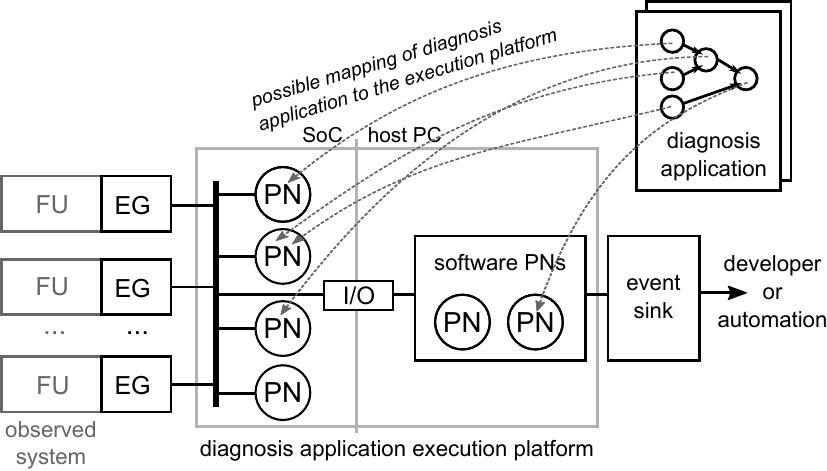}
 \caption{The architecture of our diagnosis system implementation. Event generators (EG) observe the functional units (FU). They connect via an  interconnect to different processing nodes (PN) and an off-chip interface. On a host PC processing nodes implemented in software further process the events. Finally, event sinks prepare the data for developers or automation. A diagnosis application can be mapped flexibly on the execution platform; a sample mapping is shown.}
 \label{fig:diasys_hwarch}
\end{figure}

In the previous sections we presented the diagnosis system from a functional perspective.
We continue now by presenting an implementation architecture of the diagnosis system.
It consists of extensions to the SoC, as well as software on a host PC; an exemplary architecture is shown in Figure~\ref{fig:diasys_hwarch}.

In the SoC, different functional units (FU) can be observed, like a CPU, a memory, or a bus.
Each functional unit can be attached to one or multiple event generators (EG).
The resulting events are transmitted over a diagnosis interconnect to on-chip processing nodes.
The processing nodes form the execution platform for the diagnosis application;
they will be discussed in depth in Section~\ref{sec:arch:execution_platform}.
Invisible to the diagnosis application, the execution platform spans across the chip boundary.
Through an I/O interface, a host PC is connected to the SoC.
The PC contains a software runtime environment to provide further processing nodes as part of the diagnosis application execution platform.
If all processing has been accomplished, the output events are sent to an event sink application on the host PC, which formats the output events for developers or automation.

Depending on the features and computational power provided by the processing nodes, a diagnosis application can be mapped to the execution platform in a flexible way.

\subsection{Diagnosis Application Execution Platform}
\label{sec:arch:execution_platform}
The heart of the diagnosis system are the diagnosis applications, which are executed on the diagnosis application execution platform.
As discussed before, this platform spans across the SoC and the host PC.

On the SoC, it consists of processing nodes of different types which are connected by a shared interconnect (such as a NoC or a bus).
Each processing node has an input and output interface to receive and send out events on the interconnect.
Different types of processing nodes can offer different degrees of flexibility regarding their computation.
Some might be able to perform only a single functionality specified at hardware design time, like a counter or a statistical aggregator, while others might be freely programmable.
As an example, we present in Section~\ref{sec:arch:diagnosis_processor} the Diagnosis Processor, a programmable general-purpose processing node.

As the chip area (economically) available for on-chip diagnosis processing is limited, the diagnosis application execution platform extends to the host PC.
Connected through an arbitrary off-chip interface, a runtime layer in software provides a virtually unlimited number of ``soft'' processing nodes.
Such PNs are implemented in software on the host PC and accept, like their on-chip counterpart, events as input and produce events as output.
By being executed on a host PC, they provide more compute and memory resources.

\medskip

The transformation or computation in a diagnosis application is represented by transformation actors.
For execution, they are mapped to the available processing nodes, as shown exemplary in Figure~\ref{fig:diasys_hwarch}.
An $n$:1 mapping of transformation actors to processing nodes is possible, if the combined transformation of all $n$ transformation actors can be executed by the processing node.
To achieve the greatest possible reduction in off-chip traffic, as much computation as possible should be mapped to on-chip processing nodes.
The remainder of processing is then mapped to processing nodes on a host PC, where significantly more processing power is available.

\subsubsection{The Diagnosis Processor: A Multi-Purpose Processing Node}
\label{sec:arch:diagnosis_processor}
The diagnosis processor is a freely programmable general-purpose processing node.
Like any processor design, it sacrifices computational density for flexibility.
Its design is inspired by existing scriptable debugging solutions, like SystemTap or DTrace, which have shown to provide a very useful tool for software developers in a growingly complex execution environment.
The usage scenario for this processing node are custom or one-off data analysis tasks.
This scenario is very common when searching for a bug in software.
First, a hypothesis is formed by the developer why a problem might have occurred.
Then, this hypothesis must be validated in the running system.
For this validation, a custom data analysis script must be written, which is highly specific to the problem (or the system state is manually inspected).
This process is repeated multiple times, until the root cause of the problem is found.
As this process is approached differently by every developer (and often also influenced by experience and luck), a very flexible processing node is required.

We present the hardware design of our diagnosis processor implementation in Section~\ref{sec:hwimpl:diagnosis_processor}.
We envision the programming of the diagnosis processor being done through scripts similar to the ones used by SystemTap or DTrace.
They allow to write trace analysis tasks on a similar level of abstraction as the analyzed software itself, leading to good developer productivity.

\subsection{Discussion}
The presented diagnosis system is designed to fulfill the requirements outlined in Section~\ref{sec:method:requirements}.
In the following, we discuss the consequences of the design decisions, which can limit the applicability of the diagnosis system approach in some cases.

By transforming the observed system state close to the source into denser information, the off-chip bottleneck can be circumvented.
As a downside, this lossy transformation thwarts a usage scenario of today's tracing systems.
In many of these systems, it is possible to capture a trace once, store it, and run different analysis tasks on it.
If major parts of the captured data are dismissed early, this is not possible any more.
Instead, the analysis task must be defined (as diagnosis application) before the system is run.
If the problem hypothesis changes and a different diagnosis application is required, the system must be run again.
The severity of this limitation strongly depends on how hard it is to reproduce a bug or behavior across runs.

Another feature present in many of today's tracing systems, which is explicitly not supported by the diagnosis system, are cross-triggers.
Cross-triggers are a mechanism in the tracing system to start or stop the observation, or to observe different components, based on another observation in the system.
For example, memory accesses could be traced only after a CPU executed a certain program counter.
Cross-triggers are most useful if their timing behavior is predictable.
For example, memory accesses are traced ``in the next cycle'' after the specified program counter was executed.
In GALS SoCs, such timing guarantees cannot be given; for a diagnosis application spanning across a SoC and a host PC, it is equally impossible to give (reasonably low bounded) timing guarantees.
We make this property explicit by modeling the diagnosis system as a transformational system, not a reactive system.
The commercially available tracing systems today are less specific about this.
For example, ARM CoreSight uses a handshaking protocol for cross-triggers delivered across clock boundaries, which guarantees save delivery of the signal, but does not guarantee any latency.

Instead of relying on cross-triggers to collect data from different sources at the same instant in time, we capture this data already when creating primary events through event generators.
The payload of primary events is the only way to pass multiple state observations with a defined timing relation to the diagnosis system.
For example, an event generator attached to a CPU can trigger an event based on a program counter value, and attach current contents of certain CPU registers or stack contents to it.
Using this method, it is possible to generate for example an event which informs about a function being called, and which function arguments (stored in CPU registers or on the stack) have been passed to it.
We show an example of such an event generator as part of our hardware implementation.

Finally, we discuss the system behavior in overload situations, i.e. if more input data is received than the diagnosis system can process.
Given the generally unknown input data, and the generally nondeterministic behavior of the diagnosis application, it is not possible to statically dimension the diagnosis system to be able to handle all possible input sequences.
Therefore, overload situations are unavoidable in the general (and most common) case.
If an overload situation is detected, the diagnosis system can react in multiple ways.
First, it could temporarily stall the observed system.
This gives the diagnosis system time to process outstanding events without new events being produced.
This approach is only feasible in a synchronous non-realtime system.
A more common approach is to discard incoming data until further processing resources are available.
Depending on the diagnosis application, a recovery strategy needs to be formulated.
Some applications can deal easily with incomplete input data, e.g. diagnosis applications creating statistics.
Others are not able to work with an incomplete input sequence and in consequence fail to be executed properly.

\medskip
This ends the discussion of DiaSys in general.
In the following, we present a hardware implementation of our approach, and then continue then with two usage examples how DiaSys can be put to work.

\section{Implementing DiaSys in Hardware}
\label{sec:hwimpl}

\begin{figure}
 \centering
 \scriptsize
 \includesvgnolatex{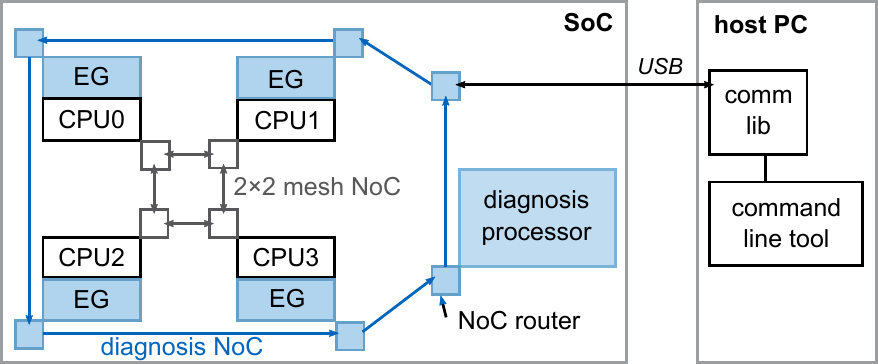}
 \caption{Block diagram of the prototype implementation. The diagnosis extensions added to the $2\times2$ multi-core system are drawn in blue.}
 \label{fig:prototype_sys}
\end{figure}

DiaSys, as presented in the previous section, can be implemented in various ways in hardware.
Our implementation, which we present in the following, is one such implementation.
It was created to answer two questions:
first, to show that DiaSys can be implemented in hardware, and second, to give resource usage numbers for one specific implementation.
As dimensioning and optimization for speed or area usage strongly depends on how DiaSys is used, a general answer to this question must remain out of scope for this work.

\medskip

The diagnosis system extends a $2 \times 2$ tiled multi-core system as shown in Figure~\ref{fig:prototype_sys}.
Our implementation runs on an FPGA and uses the OpTiMSoC framework~\cite{wallentowitz_open_2013}.
The observed system consists of four mor1kx CPU cores (an implementation of the OR1K or ``OpenRISC'' ISA), each connected to a distributed memory and a mesh NoC interconnect (components with white background).
This system is representative of the multi- and many-core architecture template currently in research and available early products, such as the Intel SCC or the Mellanox (formerly Tilera and EZchip) Tile processors.

The diagnosis system, depicted in blue, consists of the following components.
\begin{itemize}
 \item Four event generators attached to the CPUs (marked ``EG'').
 \item A single diagnosis processor.
 \item A 16~bit wide, unidirectional ring NoC, the ``diagnosis NoC,'' to connect the components of the diagnosis system. It carries both the event packets as well as the configuration and control information for the event generators and processing nodes.
 \item A USB~2.0 off-chip interface.
 \item Software support on the host PC to control the diagnosis system, and to display the results.
\end{itemize}

All components connected to the diagnosis NoC follow a common template to increase reusability.
Common parts are the NoC interface and a configuration module, which exposes readable and writable configuration registers over the NoC.
In the following, we explain the implementation of the main components in detail.

\subsection{CPU Event Generator}
\label{sec:hwimpl:implementation:eventgen_cpu}

\begin{figure}
 \centering
 \scriptsize
 \includesvgnolatex{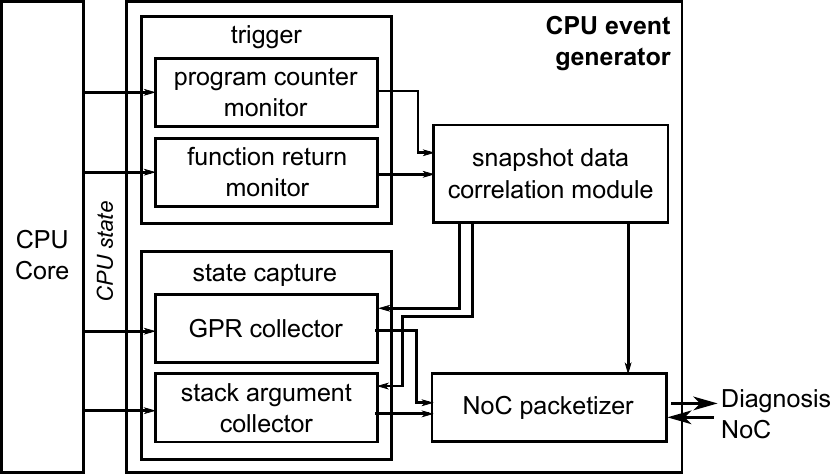}
 \caption{Block diagram of the CPU event generator.}
 \label{fig:eventgen_cpu}
\end{figure}

The CPU event generator is attached to a single CPU core.
Its main functionality is implemented in the trigger and the state capture modules.
The trigger unit of the CPU event generator triggers on two types of conditions: either the value of the program counter (PC), or the return from a function call (the jump back to the caller).
At each point in time, 12 independent trigger conditions can be monitored.
The number of monitored trigger conditions is proportional to the used hardware resources.
Our dimensioning was determined by statistical analysis of large collections of SystemTap and DTrace scripts: $\leq 9$ concurrent probes are used in 95~\% of SystemTap scripts, and $\leq 12$ concurrent probes cover 92~\% of the DTrace scripts.
The partial system state snapshot can be configured to contain the CPU register contents and the function arguments passed to the function.
A block diagram of the CPU event generator is shown in Figure~\ref{fig:eventgen_cpu}.


The PC trigger is implemented as simple comparator.
The ``function return'' trigger requires a special implementation, because no unique point in the program flow, i.e. no single PC value, describes the return from a function (a function can have multiple call sites and can return to the caller from different points in the function body). Instead, we use the following method:
\begin{enumerate}
  \item A PC trigger is set to the first instruction of the called function.
  \item If the trigger fires, the link (a.k.a. return) address is pushed to a
memory structure inside the return
    monitor, the ``return address stack.''
    The link address is the program
    counter to jump to if the function has finished its execution and the
    execution returns to the caller. On OR1K (as common on RISC
    architectures, including ARM and MIPS) the link address is stored in a CPU
    register. On other architectures and calling conventions (such as x86 and
    x86\_64), the link address is pushed to the stack.
  \item Now the system monitors the program flow for the topmost PC value in the
return address stack. If this PC is
    executed, a function returned to its caller and the function return monitor
    triggers the generation of a primary event.
\end{enumerate}

To capture the values inside CPU registers, the register writeback
signal of the mor1kx CPU is observed, and a copy of the register file is
created. This copy can then be included in the event packet if a
trigger fires.

The passing of function arguments to functions depends on the calling
convention. On OR1K, the first six data words passed to a function are
available in CPU registers, all other arguments are pushed to the stack before
calling the function. This is common for RISC architectures; other architectures
and calling conventions might pass almost all arguments on the stack (such as
x86).
To record the function arguments as part of the primary event we therefore need to  create a copy of the stack memory that can be accessed non-intrusively.
We do this by observing CPU writes to the stack memory location.

In our implementation for the mor1kx CPU we create a copy of the stack memory
by monitoring the instruction stream for a store word (\texttt{l.sw})
instruction with a target address \texttt{rA} equal to the stack pointer
\texttt{R1}. The data in the source register \texttt{(rB)}, together with a
write offset \texttt{I} (with $\texttt{I} \geq 0$, i.e. targeting the previous
stack frame) can be then used to recreate the stack frame.%
\footnote{In theory, data can be written to the stack in a different way.
However, the described way is common across compilers and used by the default
GCC compiler for OR1K.}

\subsection{Diagnosis Processor}
\label{sec:hwimpl:diagnosis_processor}

\begin{figure}
 \centering
 \includesvgnolatex{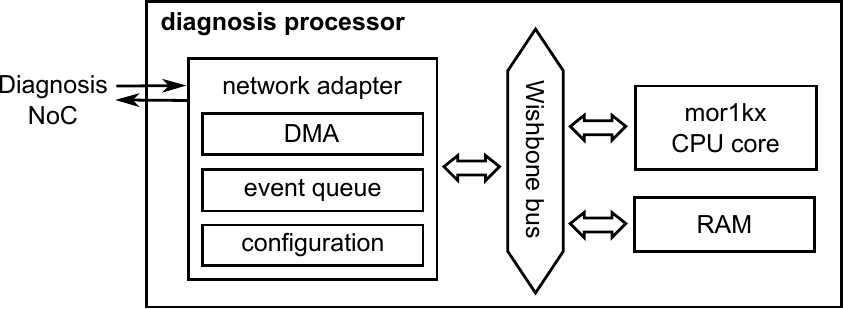}
 \caption{Block diagram of the diagnosis processor, a freely programmable processing node.}
 \label{fig:diagnosis_processor}
\end{figure}

The diagnosis processor design is extended from a standard processor template like it is used in the observed system.
The main components, shown in Figure~\ref{fig:diagnosis_processor}, are a single mor1kx CPU core and an SRAM block as program and data memory.
This system is extended with components to reduce the runtime overhead of processing event packets.

First, the network adapter, which connects the CPU to the diagnosis NoC, directly stores the incoming event packets in the memory through a DMA engine.
All event packets are processed in a run-to-completion fashion.
We can therefore avoid interrupting the CPU and instead store the address of the event to be processed next in a hardware ``run queue.''
A ``discard queue'' signals the hardware scheduler which events have been processed and can be purged from memory.

\subsection{Resource Usage}
\begin{table}
  \centering
  \footnotesize

  \begin{tabulary}{\linewidth}{lLrrr}
    \toprule
    \multicolumn{2}{l}{\textbf{Module}} &\textbf{LUTS} &  \textbf{REGS} & \textbf{RAMS}\\
    \midrule
    \multicolumn{2}{l}{observed system} & 40625 & 29638 & 80 \\
    & 1 compute tile (system contains 4)& $\sim 7232$ & $\sim$ 4763 & 20\\
    & $2\times2$ mesh NoC & 10791 & 9964 & 0\\
    & support infrastructure (DRAM if, clock/reset mgr) & 904 & 623 & 0\\
    \midrule
    \multicolumn{2}{l}{diagnosis extensions} & 19556 & 19140 & 147\\
    & 1 CPU Event Generator (fully featured)& 3603 & 6521 & 2 \\
    & 1 CPU Event Generator (reduced CoreSight-like functionality) & 1365 & 1594 & 0 \\
    & 1 Diagnosis Processor & 8614 & 4549 & 145\\
    & diagnosis NoC & 2520 & 2926 & 0\\
    \bottomrule
  \end{tabulary}

  \medskip
  \caption{The resource usage of a CPU diagnosis unit. Either the fully-featured or the reduced-functionality CPU Event Generator can be used.}
  \label{tab:resourceusage}
\end{table}

The prototype of the tiled MPSoC with the diagnosis extensions was synthesized for a ZTEX~1.15d board with a Xilinx Spartan-6 XC6SLX150 FPGA.
The relevant hardware utilization numbers as obtained from a Synplify Premier FPGA synthesis are given in Table~\ref{tab:resourceusage}.

The functional system, even though it consists of four CPU cores, is relatively small, as the used mor1kx CPU cores are lightweight (comparable to small ARM Cortex M cores).
The functional system contains no memory, but uses an external DDR memory.

In this scenario, the full diagnosis system is rather large.
We have implemented two types of CPU event generators.
A ``lite'' variant of the event generator can trigger only on a program counter value, and not on the return from a function call.
This reduced functionality makes the event generator comparable to the feature set of the ARM CoreSight ETM trace unit, which is said to use $\sim$~7,000 NAND gate equivalents~\cite{orme_debug_2008}, making it similarly sized as our event generator.
The possibility to trigger also on the return from a function call significantly increases the size of the event generator, mostly due to additional memory.
The diagnosis processor is about 20~percent larger than a regular compute tile, as it contains an additional DMA engine and the packet queues.
It also contains 30~kByte of SRAM as program and data memory, which is not present in a regular compute tile.

In summary, the resource usage of the diagnosis system is acceptable, especially if used in larger functional systems with more powerful CPU cores.
At the same time, the implementation still contains many opportunities for optimization, which we plan to explore in the future.
Also, a full system optimization to determine a suitable number of diagnosis processors and other processing nodes for a given number of CPU cores is future work.

\section{Usage Examples}
\label{sec:usage}

We designed DiaSys as general-purpose approach to gain insight into SoCs, similar to today's tracing systems.
Unfortunately, no benchmarks exist to evaluate such systems in a standardized way.
We therefore rely on two spotlight usage examples to highlight important aspects of our approach.

The first example is a hypothesis testing or ``debugging'' scenario which could also be performed using a trace-based debugger.
We included this example as a demonstration of the flexibility of our approach:
the process of debugging is usually a one-time effort, and the debugger is used as a tool to observe the program execution at various places in order to validate an hypothesis in a developer's head.

In the second example we show how to create a lock contention profile with DiaSys.
This example is taken from the area of runtime analysis, the other major area in which tracing is employed today.
While the analysis tasks in this area are more standardized (thus need a less flexible diagnosis system), they usually require the long-time observation of the whole program execution, therefore producing large data rates in todays implementations.
The creation of a lock contention profile therefore serves as a good example for the data reduction capabilities of DiaSys.


\subsection{Hypothesis Testing: Finding a Race Condition}
\label{sec:usage:racecondition}
Hypothesis testing, or simply ``debugging,'' is the most common scenario in which software developers need to get insight into the software execution.
While in many cases an intrusive debugging tool is sufficient, the more tricky bugs are related to timing, and thus require non-intrusive system insight.
Today, developers use trace-based debugging for this task.
In this example we show that DiaSys is equally suitable for such a scenario.
In the following we discuss the debugging of a race condition which occurs in an application distributed over three compute tiles.
We implement this example on our hardware implementation of DiaSys, as discussed in Section~\ref{sec:hwimpl}.

\subsubsection{Problem Description}
The application in this usage example consists of three tasks, running on three different processors concurrently.
Core 0 runs the task \texttt{bank}, which is holds a variable \texttt{balance}.
The other two cores 1 and 2 run the tasks \texttt{atm0} and \texttt{atm1}, respectively.
All communication is handled through message passing.
A message \texttt{get\_balance} reads the value of \texttt{balance} from core 0, and \texttt{set\_balance} writes it back.
The tasks \texttt{atm0} and \texttt{atm1} periodically wait for a random amount of time, then get \texttt{balance}, decrement it by 1, and write it back.

When running the application, we notice that sometimes $n$ calls to \texttt{set\_balance} do not, as expected, decrement \texttt{balance} by $n$, but by $m < n$.

\subsubsection{Debugging Approach I: Observe exchanged messages}

\begin{figure}
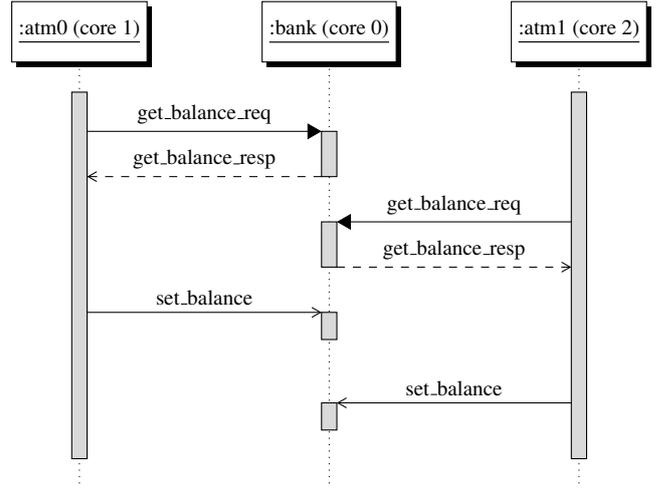

 \footnotesize
 \begin{sequencediagram}
  \newthread{atm1}{:atm0 (core 1)}
  \newinst[1.5]{bank}{:bank (core 0)}

  \newinst[1.5]{atm2}{:atm1 (core 2)}
  \stepcounter{threadnum}
  \node[below of=inst\theinstnum,node distance=0.8cm] (thread\thethreadnum) {};
  \tikzstyle{threadcolor\thethreadnum}=[fill=gray!30]
  \tikzstyle{instcoloratm2}=[fill=gray!30]

  \begin{call}{atm1}{\shortstack{get\_balance\_req}}{bank}{\shortstack{get\_balance\_resp}}
  \end{call}
  \begin{call}{atm2}{\shortstack{get\_balance\_req}}{bank}{\shortstack{get\_balance\_resp}}
  \end{call}
  \begin{messcall}{atm1}{\shortstack{set\_balance}}{bank}
  \end{messcall}
  \begin{messcall}{atm2}{\shortstack{set\_balance}}{bank}
  \end{messcall}
 \end{sequencediagram}
 \caption{Sequence diagram showing the race condition in the first case study.}
 \label{fig:evaluation:racecondition_msc}
\end{figure}

Initially, we don't know where the problem is located.
However, we assume that something in the exchange of messages goes wrong.
We therefore use DiaSys to print out all incoming and outgoing messages at task \texttt{bank} for manual analysis.

We start by configuring the CPU event generator at core 0 (running the task \texttt{bank}) to generate types of primary events:
\begin{enumerate}
 \item One event if the message passing send function is called.
   As payload we capture the identifier of the destination core and the type of the message (e.g. \texttt{get\_balance\_resp})
 \item Another event if the message passing receive function is called.
   As payload we capture the identifier of the source core and the type of the message.
\end{enumerate}

To create an event log, all events are sent directly to the host PC, where they are displayed to the developer in the form of a text log.
For the purpose of easier understanding in this paper, Figure~\ref{fig:evaluation:racecondition_msc} presents the interesting section of this text log in the form of a sequence diagram.
Looking at this diagram, experienced developers will notice the bug:
two read-modify-write sequences are interleaved, causing the value written by \texttt{atm0} to be overwritten by \texttt{atm1}.
Such behavior is a textbook example of a race condition.


\subsubsection{Debugging Approach II: Transaction checking}
The first debugging approach used DiaSys only to gather data, not to analyze it.
The less often the race condition occurs, the more trace data is generated in the first approach, which must be transferred and manually checked.
In our second approach, we also automate this checking.
As result, the developer is only informed if an actual race condition occurred.

In the correct scenario, the sequence of getting the balance, modifying it, and writing it back is an atomic transaction.
We therefore form the hypothesis ``the read-modify-write sequence is atomic,'' and use DiaSys to check it.
If the hypothesis does not hold, we have found a race condition.

We first configure the event generator at core 0 (running the task \texttt{bank}) to create two primary events:
\begin{itemize}
 \item An event \texttt{EV\_GET\_BALANCE\_CALL} when entering the function \texttt{get\_balance()} on core 0.
   This function is called if a \texttt{get\_balance\_req} message is received.
 \item Another event \texttt{EV\_SET\_BALANCE\_RETURN} when returning from the function \texttt{set\_balance()}.
   This function returns when the \texttt{set\_balance} message has been fully processed.
\end{itemize}
For both events, the source of the message (i.e. \texttt{atm0} or \texttt{atm1}) is included as payload.

Furthermore, we program the diagnosis processor to execute a transformation actor shown in pseudo code in Listing~\ref{lst:tn_check_transaction}.
(In our hardware implementation, we programmed the diagnosis processor in C with an equivalent program.)

\begin{lstlisting}[label=lst:tn_check_transaction,
  caption={Pseudo code of the transformation actor checking the balance updating transaction.},float]
TA_CHECK_BALANCE_TRANS {
  bool in_transaction = false;
  bool transaction_owner = NULL;

  event = wait(EV_GET_BALANCE_CALL
               or EV_SET_BALANCE_RETURN);

  // get_balance(src) and set_balance(src)
  // are both passed the source of the request
  // message as first function argument
  msg_src = event.data.args['src'];

  if (in_transacation
      && transaction_owner != msg_src) {
    // race condition found
    event_type = EV_RACE_DETECTED;
    event_data = {};
    return new Event(event_type, event_data);
  }

  if (event.type == EV_GET_BALANCE_CALL) {
    // start of new transaction
    in_transacation = true;
    transaction_owner = msg_src;
    return;
  }
  if (event.type == EV_SET_BALANCE_RETURN
      && in_transacation
      && transaction_owner != msg_src) {
    // race condition found
    event_type = EV_RACE_DETECTED;
    event_data = {};
    return new Event(event_type, event_data);
  }
}
\end{lstlisting}

The two primary events are now routed from core 0 to the diagnosis processor, which checks the hypothesis that all transactions are atomic.
If a violation, i.e. a race condition, is found, a new event \texttt{EV\_RACE\_DETECTED} is generated.
This event is sent to the host PC and displayed as an event log to the developer.

\subsubsection{Event and Data Rates}
The most significant benefit from automating the hypothesis checking, as shown in the previous section, is the increase in productivity for the developer: even in long-running programs, testing the hypothesis becomes straightforward.
At the same time, the automation also reduces the required trace bandwidth.

The first approach of creating events for all received and transmitted messages for manual inspection required a trace bandwidth of 57.09~KBit/s.
When automating the checking in the second approach, the bandwidth was reduced to 5.66~KBit/s, a reduction by $10.1\times$.

The bandwidth requirements are heavily dependent on the program characteristics.
In the program executions on our hardware platform running at 50~MHz, we observed 44~\% of transactions were interleaved, i.e. a race condition.

Doing the same analysis in a tracing system would require an instruction and a data trace to be captured.
Even though our evaluation platform is not able to produce such traces, we can estimate the required trace bandwidth by measuring the number of executed instructions and data memory accesses.
Assuming an instruction trace compressed to 2~bit/instruction and a data trace compressed to 16 bit/access (for data and address, following \cite{hopkins_debug_2006}), we get a total trace stream of 246.31~MBit/s.

It must be noted, however, that this number is an upper bound on the required bandwidth for a tracing system.
Depending on the used tracing system implementation (such as CoreSight), not a full instruction or data trace is captured, but filters and triggers can be used to reduce this stream to just relevant parts, similar to what our event generators do.

\subsubsection{Outlook: Programming DiaSys}
In our first case study, we have shown how to use DiaSys to obtain data for manual analysis from the SoC, and how to automate the manual checking as hypothesis test running on-chip on the diagnosis processor.

In this example, we wrote the transformation actor manually as C program, which is called upon receiving events.
In the future, we envision also other means to describe the diagnosis application executed by DiaSys.
In this example, a description of the desired system behavior as linear temporal logic (LTL) expression could be applied, as it commonly done in runtime validation systems.
Using such LTL expressions for the validation of the correct hardware configuration by the software has been presented in \cite{li_rule-based_2016}.

\subsection{Runtime Analysis: Creating a Lock Contention Profile}
\label{sec:usage:lockprofiling}
In our second usage example, we present a diagnosis application which generates a ``profile,'' a statistic about some software behavior.
Examples for such commonly used profiles are the various \texttt{*top} Linux commands which present an ordered list of processes or threads with the highest usage of CPU, I/O, memory, or other resources.
We include this usage example as it represents, next to debugging, the second large motivation to use a tracing tool today: runtime analysis.
Other examples from this category of applications are code coverage and runtime profiling.
Most runtime analysis tasks today require a full (instruction) trace to be recorded, which requires a high off-chip bandwidth.
Using DiaSys, we can show how this trace bandwidth can be reduced by on-chip processing.

To give meaningful insight into data rates generated when analyzing large real-life applications, we use in this example not a self-created example application, but an application from the PARSEC benchmark suite.


\subsubsection{Evaluation Prototype}
\label{sec:usage:lockprofiling:software_prototype}
\begin{figure}
 \includesvgnolatex{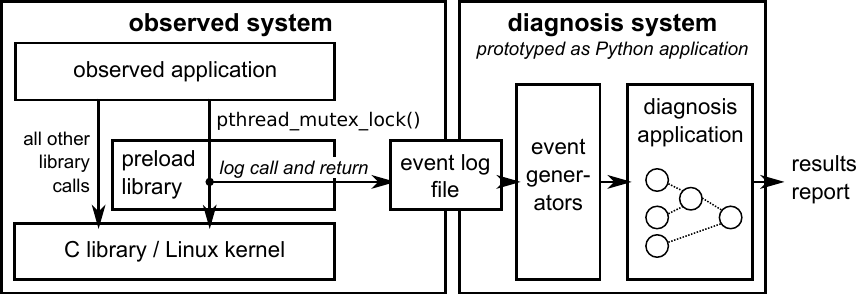}
 \caption{The software prototype of the diagnosis system. All calls in the observed application are recorded in an event log file by an preloaded library. The event log file is read by the diagnosis system implemented as Python application. The figure shows the monitoring of all \texttt{pthread\_mutex\_lock()} calls and returns as used by the usage example in Section~\ref{sec:usage:lockprofiling}.}
 \label{fig:software_prototype}
\end{figure}

The hardware implementation prototype of DiaSys presented in Section~\ref{sec:hwimpl} is only able to run a very limited set of applications, as it only provides bare-metal programming support (comparable to a microcontroller without operating system), and no full POSIX environment as it can be found on Linux for example.
To run larger applications, such as standard benchmarks, we therefore created a software prototype of the diagnosis system.
It runs purely in software on a Linux PC and is best suited for an evaluation of event rates inside the diagnosis system, as well as the design of new diagnosis applications.
Since no hardware extensions are used, its operation is intrusive, i.e. the timing of the observed application is slightly changed.
The prototypical event generators can only trigger on the call of and return from a C library function, and the function arguments can be included in the event as data items.

The software prototype consists of two parts, which are shown in Figure~\ref{fig:software_prototype}.
The first part is a ``preload library.''
It is a small software library written in C which is able to monitor all calls to C library functions and write them into an event log file.
This event log file is then used by a prototype of the diagnosis system implemented in Python.
It consists of event generators, which read the event log file.
A set of Python functions connected by channel objects represent the transformation actors.
(We assume a one-to-one mapping of transformation actors to processing nodes in this prototype.)
The output of the diagnosis application is directly printed to a console, as specialized event sinks are not necessary for our evaluations.

We now use this software prototype to create the lock contention profile.

\subsubsection{Problem Description}
A \emph{lock contention} occurs in concurrent programs if multiple threads try to acquire a mutex lock at the same time~\cite[p. 147]{herlihy_art_2008}.
In this case, all but one threads have to wait for the lock to be released before they can continue processing.
Therefore, the lock acquisition time is a good metric for program efficiency: the less time it takes, the earlier the thread is done with its work.

In order for a developer to get insight into the lock contention behavior of the program, a contention profile can be created.
It lists all acquired locks, together with the summarized and averaged times the acquisition took.
Such a profile can be generated in an intrusive way with tools like Intel VTune Amplifier or mutrace\footnote{\url{http://0pointer.de/blog/projects/mutrace.html}}, and is traditionally formatted as shown in Listing~\ref{lst:lockprofile_output}.

\subsubsection{Measurement Approach}
The lock acquisition time can be measured by obtaining the time the mutex lock function took to execute.
In applications using pthreads, as it is the case for almost all applications running on Linux, macOS or BSD, the mutex lock function is named \verb|pthread_mutex_lock()|.

\begin{lstlisting}[label=lst:mutex_lock,
  caption={A sketch of the \texttt{pthread\_mutex\_lock()} function. This function must be executed atomically, i.e. without interruption.}]
int pthread_mutex_lock(pthread_mutex_t *mutex) {
  blocking_wait_until_mutex_is_free(mutex);
  lock_mutex(mutex);
  return 0 /* success */;
}
\end{lstlisting}

As shown in the simplified code sketch in Listing~\ref{lst:mutex_lock}, the function blocks for an indefinite amount of time until a lock is available.
If it is available, it acquires the lock and returns.

To create a lock contention profile, we need to measure the execution times of all \verb|pthread_mutex_lock()| function calls in all threads.
We then group this measurement by lock, given by the argument \verb|mutex| of the lock function, to obtain the number of times a lock was acquired, how long all lock acquisitions took in summary, and on average.

\subsubsection{Diagnosis Application}
\begin{figure}
 \includesvgnolatex{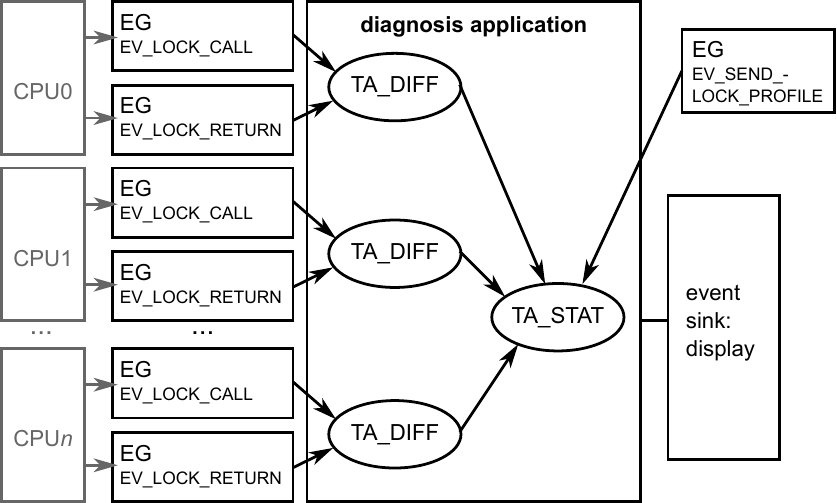}
 \caption{A diagnosis application to create a lock contention profile.}
 \label{fig:evaluation_lockprofile_appmodel}
\end{figure}

To perform the analysis outlined in the previous section, we configure the diagnosis system as shown in Figure~\ref{fig:evaluation_lockprofile_appmodel}.
First, we attach two event generators to each CPU in the observed system.
We configure them to generate two primary events which together measure the execution time of the lock acquisition function \texttt{pthread\_mutex\_lock()}.
\begin{itemize}
 \item One primary event \texttt{EV\_LOCK\_CALL} is triggered if the CPU \emph{enters (calls)} the \texttt{pthread\_mutex\_lock()} function.
   The first function argument to \texttt{pthread\_\-mutex\_\-lock}, the \texttt{mutex}, is attached to the event as data item, together with a timestamp containing the current time.
 \item Another primary event \texttt{EV\_LOCK\_RETURN} is triggered if the CPU \emph{returns} from the \texttt{pthread\_mutex\_lock()} function.
   For this event, only a timestamp is attached as event data.
\end{itemize}
To calculate the execution time of the function \texttt{pthread\_\-mutex\_\-lock()}, we create a transformation actor \texttt{TA\_DIFF}, the pseudo code of which is shown in Listing~\ref{lst:tn_time_diff}.

\begin{lstlisting}[label=lst:tn_time_diff,
  caption={Pseudo code of the transformation actor calculating the lock acquisition time.},float]
TA_DIFF {
  lock_call = wait(EV_LOCK_CALL);
  lock_return = wait(EV_LOCK_RETURN);

  uint16_t time = lock_return.data.timestamp
                  - lock_call.data.timestamp;
  uint16_t mutex_hash
    = hash(lock_call.data.args['mutex']);
  event_type = lock_acq_time;
  event_data = {
    time: time,
    lock: mutex_hash
  };
  return new Event(event_type, event_data);
}
\end{lstlisting}

It waits for both primary events \texttt{EV\_\-LOCK\_\-CALL} and \texttt{EV\_\-LOCK\_\-RETURN}, calculates the difference between the timestamps, and creates a new event \texttt{EV\_LOCK\_ACQ\_TIME} with two data items, the lock acquisition time and a hash of the \texttt{mutex} argument to reduce the data size.

As last step in the processing, all \texttt{EV\_LOCK\_ACQ\_TIME} events are aggregated by another transformation actor called \texttt{TA\_STAT}.
Again, a pseudo code implementation is given in Listing~\ref{lst:tn_stat}.

\begin{lstlisting}[label=lst:tn_stat,
  caption={Pseudo code showing the functionality of the transformation actor creating a lock profile.},float]
TA_STAT {
  event = wait(EV_LOCK_ACQ_TIME
               or EV_SEND_LOCK_PROFILE);

  // aggregate
  if (event.type == EV_LOCK_ACQ_TIME) {
    m = event.data.mutex;
    stat[m]['cnt']++;
    stat[m]['t_sum'] += event.data.time;
    return;
  }

  // send statistics output to host PC
  if (event.type == EV_SEND_LOCK_PROFILE) {
    event_type = EV_LOCK_PROFILE;
    event_data = {stat: stat};
    return new Event(event_type, event_data);
  }
}
\end{lstlisting}

If an event of type \texttt{EV\_LOCK\_ACQ\_TIME} is received, the timestamp is added to a hash data structure which records, grouped by the mutex, the number of calls to the lock function and the total time these calls took.
After the program run, on request of the developer running the diagnosis, or in regular time intervals, a \texttt{EV\_SEND\_LOCK\_PROFILE} is generated.
If this event is received, the aggregated statistics are sent to the event sink, which then presents the aggregated results to the developer.

\subsubsection{Evaluation of the Diagnosis Application}
In the evaluation of this usage example we focus on the event and data rates between the event generators and transformation nodes.
In order to provide realistic inputs, we profiled the \emph{dedup} application from the PARSEC~3.0 Benchmark Suite with the large input data sets~\cite{bienia11benchmarking}.
As PARSEC does not run on our custom-built prototype MPSoC platform, we used the software prototype described in Section~\ref{sec:usage:lockprofiling:software_prototype}.
All transformation actors were implemented in Python code equivalent to the pseudo code listings presented in Listings~\ref{lst:tn_time_diff} and~\ref{lst:tn_stat}.

\paragraph{Output of the Diagnosis Application}
Before we analyze the diagnosis application itself, we discuss the output it generates, i.e. the lock contention profile shown in Listing~\ref{lst:lockprofile_output}.
PARSEC was instructed to use at least 4 threads;
ultimately 16 threads were spawned by the dedup application. (There is no option in PARSEC to specify the exact number of threads used.)
The execution of the observed application took 2.68~s.

\begin{lstlisting}[label=lst:lockprofile_output,
  caption={Output of the lock contention profile diagnosis application observing the PARSEC dedup application.}]
     mutex           # acq.    sum [ns]  avg [ns]
(01) 0x7fd9ac018988   47785     8835387    184.90
(02) 0x7fd9d1ed2978   47784   226012031   4729.87
(03)      0x1c36500    9426    53724035   5699.56
(04)      0x1c36660    9423    21904608   2324.59
(05)      0x1c36710    4638    12528702   2701.32
(06)      0x1c365b0     105       46999    447.61
(07) 0x7fd9d2091430       8        1974    246.75
(08) 0x7fd9b41948f8       8        2277    284.62
(09) 0x7fd9b42b9ad8       8        2560    320.00
(10) 0x7fd9d20f8928       8        2215    276.88
\end{lstlisting}

The output shows the top ten most acquired mutexes, together with the total and averaged lock acquisition time.
Notable in this profile are mutexes 2 to 5, which take on average significantly longer to acquire: these locks are called to be ``contended.''

A profile helps to understand the program behavior and serves as a starting point to fix possible bugs or inefficiencies.
If lock contention is observed (and performance goals of the application are not met), it is common to replace coarse-grained locks with more fine-grained locks, i.e. locks which protect a shorter critical section.
However, fixing a bug is not in the scope of this work.
Instead, we now turn our discussion to the event and data rates generated when executing the diagnosis application that generated the profile as shown.

\paragraph{Event and Data Rates}
We designed the diagnosis system to reduce the off-chip traffic by moving the data analysis partially into the SoC.
To evaluate if the data rates are in fact reduced, we analyze event rates within the diagnosis application.

We use the following event sizes:
\begin{itemize}
 \item An \texttt{EV\_LOCK\_CALL} event requires 14 bytes: two bytes for the event type identifier, four bytes for the timestamp, and eight bytes for the \texttt{mutex} argument.
 \item An \texttt{EV\_LOCK\_RETURN} event requires six bytes: two bytes for the event type identifier and four bytes for the timestamp.
 \item An \texttt{EV\_LOCK\_ACQ\_TIME} event requires six bytes: two bytes for the event type identifier, two bytes for the lock acquisition time, and two bytes for the hashed \texttt{mutex} argument.
\end{itemize}

Over the whole program run, the event generators attached to the 16 CPUs generate a total of 516,254 events, which equals 4.9~MByte of transmitted data or, over the program runtime, an average data rate of 14.7~MBit/s.
The \texttt{TA\_DIFF} transformation actors half the number of events, resulting in a data rate of 4.4~MBit/s, or a reduction to 30~\%.
Finally, after being aggregated by \texttt{TA\_STAT}, the full result can be transferred off-chip with roughly 204~bytes.

\medskip

A direct comparison of our results to existing tracing systems is challenging.
For our analysis we need access to the \texttt{mutex} function argument through a data trace, which is not supported by higher-speed tracing implementations such as CoreSight PTM and Intel PT.
However, as a first lower-bound estimation of the data rate generated by a state-of-the-art tracing system, we created a full instruction trace using Intel PT.
The same PARSEC dedup application created a trace file of 1.82~GB, which corresponds to 5.4~GBit/s over the program runtime.

In summary, DiaSys was able to reduce the required trace bandwidth compared to an Intel PT instruction trace significantly due to on-chip analysis.
When transferring data off-chip after processing in the \texttt{TA\_DIFF} processing nodes, the bandwidth is reduced from more than 5.4~GBit/s to 4.4~MBit/s, a reduction by $1233 \times$.

\paragraph{Discussion}
Depending on the feature set and timestamp granularity of the various tracing implementations, the bandwidth reduction that DiaSys is able to achieve can vary.
However, a general observation holds: the most significant bandwidth savings result from the fact that we very precisely capture only data in the event generators which is relevant to our problem.
The subsequent processing step \texttt{TA\_DIFF} of calculating the time difference between two events is further able to discard roughly $\nicefrac{2}{3}$ of the data.
Both operations are simple enough to be implemented even in resource-constraint on-chip environments.
The final step \texttt{TA\_STAT} is again able to give large percentage-wise reductions in data rate, however the absolute savings might not justify an on-chip processing any more.
This last step could therefore be executed on the host PC -- without changing the diagnosis application.

%


\section{Conclusions}
\label{sec:conclusion}
In this paper we introduced DiaSys, a diagnosis system which aims to replace tracing systems in MPSoCs, where software observability is limited by the off-chip bottleneck.
To avoid this bottleneck, we move parts of the data analysis into the chip and closer to the data source.
The diagnosis system consists of event generators, which observe the functional units executing software on the SoC, the diagnosis application and event sinks.
Diagnosis applications describe the data analysis task in a way that is understandable for the developer and portable across different SoCs.
In detail we discussed their properties and semantics.
Diagnosis applications are portable by design, because components of the application can be freely mapped to distributed diagnosis extensions inside the SoC, or to a runtime environment on the host PC.
The implementation of such a mapping tool is future work.

In our evaluation we showed on two prototypes that the implementation of the required diagnosis extensions is feasible with reasonable hardware overhead.
We also showed in two usage examples from different domains that the envisioned reduction in off-chip bandwidth requirements can be achieved.

In the future, we plan to extend this system with more specialized processing nodes, which are suited for common analysis tasks.
We also investigate how machine-learning approaches can be used to dynamically adjust the analysis tasks during runtime.

\subsection*{Acknowledgments}
This work was funded by the Bayerisches Staatsministerium für Wirtschaft und Medien, Energie und Technologie (StMWi) as part of the project ``SoC Doctor,''
and by the German Research Foundation (DFG) as part of the Transregional Collaborative Research Centre ``Invasive Computing'' (SFB/TR 89).
The responsibility for the content remains with the authors.
We would like to thank Ignacio Alonso and Markus Göhrle for their contributions to the implementation of the evaluation platform. We especially thank Stefan Wallentowitz for the creation and ongoing support of OpTiMSoC.

\bibliographystyle{elsarticle-num}
\bibliography{references}

\begin{thebibliography}{10}
\expandafter\ifx\csname url\endcsname\relax
  \def\url#1{\texttt{#1}}\fi
\expandafter\ifx\csname urlprefix\endcsname\relax\def\urlprefix{URL }\fi
\expandafter\ifx\csname href\endcsname\relax
  \def\href#1#2{#2} \def\path#1{#1}\fi

\bibitem{vermeulen_debugging_2014}
B.~Vermeulen, K.~Goossens, Debugging {{Systems}}-on-{{Chip}}:
  {{Communication}}-Centric and {{Abstraction}}-Based {{Techniques}},
  {Springer}, New York, 2014.

\bibitem{CoreSightWebsite}
{{CoreSight}} - {{ARM}},
  \url{http://www.arm.com/products/system-ip/coresight-debug-trace}.

\bibitem{_nexus_2003}
The {{Nexus}} 5001 {{Forum Standard}} for a {{Global Embedded Processor Debug
  Interface}}, {{Version}} 2.0, Tech. rep. (Dec. 2003).

\bibitem{ipextreme_infineon_2008}
{IPextreme}, Infineon {{Multi}}-{{Core Debug Solution}}: {{Product Brochure}}
  (2008).

\bibitem{_intel_2015}
{{Intel}}\textregistered{} {{Trace Hub Developer}}'s {{Manual}}, revision 1.0
  Edition, 2015.

\bibitem{hopkins_debug_2006}
A.~B.~T. Hopkins, K.~D. McDonald-Maier, Debug support strategy for
  systems-on-chips with multiple processor cores, IEEE Transactions on
  Computers 55~(2) (2006) 174 -- 184.
\newblock \href {http://dx.doi.org/10.1109/TC.2006.22}
  {\path{doi:10.1109/TC.2006.22}}.

\bibitem{orme_debug_2008}
W.~Orme, Debug and {{Trace}} for {{Multicore SoCs}} ({{ARM}} white paper) (Sep.
  2008).

\bibitem{uzelac_real-time_2010}
V.~Uzelac, A.~Milenkovi{\'c}, M.~Burtscher, M.~Milenkovi{\'c}, Real-time
  unobtrusive program execution trace compression using branch predictor
  events, in: Proceedings of the 2010 {{International Conference}} on
  {{Compilers}}, {{Architectures}} and {{Synthesis}} for {{Embedded Systems}},
  CASES '10, {ACM}, New York, NY, USA, 2010, pp. 97--106.
\newblock \href {http://dx.doi.org/10.1145/1878921.1878938}
  {\path{doi:10.1145/1878921.1878938}}.

\bibitem{cantrill_dynamic_2004}
B.~M. Cantrill, M.~W. Shapiro, A.~H. Leventhal, Dynamic {{Instrumentation}} of
  {{Production Systems}}, in: Proceedings of the {{General Track}}: 2004
  {{USENIX Annual Technical Conference}}, ATEC '04, {USENIX Association},
  Berkeley, CA, USA, 2004.

\bibitem{eigler_architecture_2005}
F.~C. Eigler, V.~Prasad, W.~Cohen, H.~Nguyen, M.~Hunt, J.~Keniston, B.~Chen,
  Architecture of systemtap: A {{Linux}} trace/probe tool, 2005.

\bibitem{olsson_dataflow_1991}
R.~A. Olsson, R.~H. Crawford, W.~W. Ho, A {{Dataflow Approach}} to
  {{Event}}-based {{Debugging}}, Software\textemdash{}Practice \& Experience
  21~(2) (1991) 209--229.
\newblock \href {http://dx.doi.org/10.1002/spe.4380210207}
  {\path{doi:10.1002/spe.4380210207}}.

\bibitem{marceau_dataflow_2004}
G.~Marceau, G.~Cooper, S.~Krishnamurthi, S.~Reiss, A dataflow language for
  scriptable debugging, in: 19th {{International Conference}} on {{Automated
  Software Engineering}}, 2004. {{Proceedings}}, 2004, pp. 218--227.
\newblock \href {http://dx.doi.org/10.1109/ASE.2004.1342739}
  {\path{doi:10.1109/ASE.2004.1342739}}.

\bibitem{ducasse_coca_1999}
M.~Ducass{\'e}, Coca: {{An Automated Debugger}} for {{C}}, in: Proceedings of
  the 21st {{International Conference}} on {{Software Engineering}}, ICSE '99,
  {ACM}, New York, NY, USA, 1999, pp. 504--513.
\newblock \href {http://dx.doi.org/10.1145/302405.302682}
  {\path{doi:10.1145/302405.302682}}.

\bibitem{lumpp_specification_1990}
J.~Lumpp, T.~Casavant, H.~Siegel, D.~Marinescu, Specification and
  identification of events for debugging and performance monitoring of
  distributed multiprocessor systems, in: Proceedings of the 10th
  {{International Conference}} on {{Distributed Computing Systems}}, 1990, pp.
  476--483.
\newblock \href {http://dx.doi.org/10.1109/ICDCS.1990.89317}
  {\path{doi:10.1109/ICDCS.1990.89317}}.

\bibitem{harris_digital_2012}
D.~Harris, S.~Harris, {Digital Design and Computer Architecture}, 2nd Edition,
  {Elsevier Ltd, Oxford}, Amsterdam, 2012.

\bibitem{fidge_fundamentals_1996}
C.~Fidge, Fundamentals of distributed system observation, IEEE Software 13~(6)
  (1996) 77--83.
\newblock \href {http://dx.doi.org/10.1109/52.542297}
  {\path{doi:10.1109/52.542297}}.

\bibitem{gray_why_1986}
J.~Gray, Why do computers stop and what can be done about it?, in: Symposium on
  Reliability in Distributed Software and Database Systems, {Los Angeles, CA,
  USA}, 1986, pp. 3--12.

\bibitem{halbwachs_synchronous_1991}
N.~Halbwachs, P.~Caspi, P.~Raymond, D.~Pilaud, The synchronous data flow
  programming language {{LUSTRE}}, Proceedings of the IEEE 79~(9) (1991)
  1305--1320.

\bibitem{kahn_semantics_1974}
G.~Kahn, The {{Semantics}} of {{Simple Language}} for {{Parallel Programming}},
  in: {{IFIP Congress}}, 1974, pp. 471--475.

\bibitem{lee_dataflow_1995}
E.~A. Lee, T.~M. Parks, Dataflow process networks, Proceedings of the IEEE
  83~(5) (1995) 773--801.
\newblock \href {http://dx.doi.org/10.1109/5.381846}
  {\path{doi:10.1109/5.381846}}.

\bibitem{wallentowitz_open_2013}
S.~Wallentowitz, P.~Wagner, M.~Tempelmeier, T.~Wild, A.~Herkersdorf, Open
  {{Tiled Manycore System}}-on-{{Chip}}, arXiv:1304.5081 [cs]\href
  {http://arxiv.org/abs/1304.5081} {\path{arXiv:1304.5081}}.

\bibitem{li_rule-based_2016}
L.~Li, P.~Wagner, R.~Ramaswamy, A.~Mayer, T.~Wild, A.~Herkersdorf, A
  {{Rule}}-based {{Methodology}} for {{Hardware Configuration Validation}} in
  {{Embedded Systems}}, in: Proceedings of the 19th {{International Workshop}}
  on {{Software}} and {{Compilers}} for {{Embedded Systems}} ({{SCOPES}} 2016),
  Sankt Goar, Germany, 2016.

\bibitem{herlihy_art_2008}
M.~Herlihy, N.~Shavit, The {{Art}} of {{Multiprocessor Programming}}, {Morgan
  Kaufmann Publishers Inc.}, San Francisco, CA, USA, 2008.

\bibitem{bienia11benchmarking}
C.~Bienia, Benchmarking {{Modern Multiprocessors}}, Ph.D. thesis, Princeton
  University (Jan. 2011).

\end{thebibliography}

\end{document}